
\input epsf
\newfam\scrfam
\batchmode\font\tenscr=rsfs10 \errorstopmode
\ifx\tenscr\nullfont
        \message{rsfs script font not available. Replacing with calligraphic.}
        
\else   
        \font\sevenscr=rsfs7
        \font\fivescr=rsfs5
        \skewchar\tenscr='177 \skewchar\sevenscr='177 \skewchar\fivescr='177
        \textfont\scrfam=\tenscr \scriptfont\scrfam=\sevenscr
        \scriptscriptfont\scrfam=\fivescr

\fi
\catcode`\@=11
\newfam\frakfam
\batchmode\font\tenfrak=eufm10 \errorstopmode
\ifx\tenfrak\nullfont
        \message{eufm font not available. Replacing with italic.}
        
\else
	
	\font\sevenfrak=eufm7 \font\fivefrak=eufm5
	\textfont\frakfam=\tenfrak
	\scriptfont\frakfam=\sevenfrak \scriptscriptfont\frakfam=\fivefrak
	
\fi
\catcode`\@=\active
\newfam\msbfam
\batchmode\font\twelvemsb=msbm10 scaled\magstep1 \errorstopmode
\ifx\twelvemsb\nullfont\def\Bbb{\bf}

	\message{Blackboard bold not available. Replacing with boldface.}
\else   \catcode`\@=11
        \font\tenmsb=msbm10 \font\sevenmsb=msbm7 \font\fivemsb=msbm5
        \textfont\msbfam=\tenmsb
        \scriptfont\msbfam=\sevenmsb \scriptscriptfont\msbfam=\fivemsb
        \def\Bbb{\relax\expandafter\Bbb@}
        \def\Bbb@#1{{\Bbb@@{#1}}}
        \def\Bbb@@#1{\fam\msbfam\relax#1}
        \catcode`\@=\active

\fi
        \font\fivemi=cmmi5
        \font\sixmi=cmmi6
        \font\eightrm=cmr8              \def\xrm{\eightrm}
        \font\eightbf=cmbx8             \def\xbf{\eightbf}
        \font\eightit=cmti10 at 8pt     \def\xit{\eightit}
                
        \font\eighttt=cmtt8             
        \font\eightcp=cmcsc8
        \font\eighti=cmmi8              \def\xold{\eighti}
        \font\eightmi=cmmi8
        \font\eightib=cmmib8             \def\xbold{\eightib}
        \font\teni=cmmi10               \def\old{\teni}
        \font\tencp=cmcsc10

        \font\twelvecp=cmcsc10 scaled\magstep1
        
        \font\sixrm=cmr6
        \font\fiverm=cmr5

        \font\eightsy=cmsy8
        \font\sixsy=cmsy6
        \font\eightsl=cmsl8
        \font\sixbf=cmbx6

	 at10pt	
	\font\twelvehelvbold=phvb at12pt
	 at14pt
	\font\sixteenhelvbold=phvb at16pt

\def\noblackbox{\overfullrule=0pt}
\noblackbox

\def\eightpoint{
\def\rm{\fam0\eightrm}
\textfont0=\eightrm \scriptfont0=\sixrm \scriptscriptfont0=\fiverm
\textfont1=\eightmi  \scriptfont1=\sixmi  \scriptscriptfont1=\fivemi
\textfont2=\eightsy \scriptfont2=\sixsy \scriptscriptfont2=\fivesy
\textfont3=\tenex   \scriptfont3=\tenex \scriptscriptfont3=\tenex
\textfont\itfam=\eightit \def\it{\fam\itfam\eightit}
\textfont\slfam=\eightsl \def\sl{\fam\slfam\eightsl}
\textfont\ttfam=\eighttt \def\tt{\fam\ttfam\eighttt}
\textfont\bffam=\eightbf \scriptfont\bffam=\sixbf 
                         \scriptscriptfont\bffam=\fivebf
                         \def\bf{\fam\bffam\eightbf}
\normalbaselineskip=10pt}

\newtoks\headtext
\headline={\ifnum\pageno=1\hfill\else
	\ifodd\pageno{\eightcp\the\headtext}{ }\dotfill{ }{\old\folio}
	\else{\old\folio}{ }\dotfill{ }{\eightcp\the\headtext}\fi
	\fi}
\def\makeheadline{\vbox to 0pt{\vss\noindent\the\headline\break
\hbox to\hsize{\hfill}}
        \vskip2\baselineskip}
\newcount\infootnote
\infootnote=0
\def\foot#1#2{\infootnote=1
\footnote{${}^{#1}$}{\vtop{\baselineskip=.75\baselineskip
\advance\hsize by
-\parindent{\eightpoint\rm\hskip-\parindent #2}\hfill\vskip\parskip}}\infootnote=0}
\newcount\refcount
\refcount=1
\newwrite\refwrite
\def\oldsize{\ifnum\infootnote=1\xold\else\old\fi}
\def\ref#1#2{
	\def#1{{{\oldsize\the\refcount}}\ifnum\the\refcount=1\immediate\openout\refwrite=\jobname.refs\fi\immediate\write\refwrite{\item{[{\xold\the\refcount}]} 
	#2\hfill\par\vskip-2pt}\xdef#1{{\noexpand\oldsize\the\refcount}}\global\advance\refcount by 1}
	}
\def\refout{\catcode`\@=11
        \xrm\immediate\closeout\refwrite
        \vskip2\baselineskip
        {\noindent\twelvecp References}\hfill\vskip\baselineskip
        \baselineskip=.75\baselineskip
        \input\jobname.refs
        \baselineskip=4\baselineskip \divide\baselineskip by 3
        \catcode`\@=\active\rm}

\def\skipref#1{\hbox to15pt{\phantom{#1}\hfill}\hskip-15pt}

\def\hepth#1{\href{http://xxx.lanl.gov/abs/hep-th/#1}{arXiv:hep-th/{\xold#1}}}

\def\arxiv#1#2{\href{http://arxiv.org/abs/#1.#2}{arXiv:{\xold#1}.{\xold#2}}}
\def\jhep#1#2#3#4{\href{http://jhep.sissa.it/stdsearch?paper=#2\%28#3\%29#4}{J. High Energy Phys. {\xbold #1#2} ({\xold#3}) {\xold#4}}}

\def\ATMP#1#2#3{Adv. Theor. Math. Phys. {\xbold#1} ({\xold#2}) {\xold#3}}

\def\CQG#1#2#3{Class. Quantum Grav. {\xbold#1} ({\xold#2}) {\xold#3}}

\def\JHEP{\jhep}

\def\LMP#1#2#3{Lett. Math. Phys. {\xbold#1} ({\xold#2}) {\xold#3}}
\def\MPLA#1#2#3{Mod. Phys. Lett. {\xbf A}{\xbold#1} ({\xold#2}) {\xold#3}}

\def\NPB#1#2#3{Nucl. Phys. {\xbf B}{\xbold#1} ({\xold#2}) {\xold#3}}

\def\PLB#1#2#3{Phys. Lett. {\xbf B}{\xbold#1} ({\xold#2}) {\xold#3}}

\def\PTPS#1#2#3{Progr. Theor. Phys. Suppl. {\xbold#1} ({\xold#2}) {\xold#3}}
\newcount\sectioncount
\sectioncount=0
\def\section#1#2{\global\eqcount=0
	\global\subsectioncount=0
        \global\advance\sectioncount by 1
	\ifnum\sectioncount>1
	        \vskip2\baselineskip
	\fi
\line{\twelvecp\the\sectioncount. #2\hfill}
       \vskip.5\baselineskip\noindent
        \xdef#1{{\old\the\sectioncount}}}
\newcount\subsectioncount
\def\subsection#1#2{\global\advance\subsectioncount by 1
\vskip.75\baselineskip\noindent\line{\tencp\the\sectioncount.\the\subsectioncount. #2\hfill}\nobreak\vskip.4\baselineskip\nobreak\noindent\xdef#1{{\old\the\sectioncount}.{\old\the\subsectioncount}}}
\def\immediatesubsection#1#2{\global\advance\subsectioncount by 1
\vskip-\baselineskip\noindent
\line{\tencp\the\sectioncount.\the\subsectioncount. #2\hfill}
	\vskip.5\baselineskip\noindent
	\xdef#1{{\old\the\sectioncount}.{\old\the\subsectioncount}}}
\newcount\appendixcount
\appendixcount=0
\def\appendix#1{\global\eqcount=0
        \global\advance\appendixcount by 1
        \vskip2\baselineskip\noindent
        \ifnum\the\appendixcount=1
        \hbox{\twelvecp Appendix A: #1\hfill}\vskip\baselineskip\noindent\fi
    \ifnum\the\appendixcount=2
        \hbox{\twelvecp Appendix B: #1\hfill}\vskip\baselineskip\noindent\fi
    \ifnum\the\appendixcount=3
        \hbox{\twelvecp Appendix C: #1\hfill}\vskip\baselineskip\noindent\fi}
\def\acknowledgements{\vskip2\baselineskip\noindent
        \underbar{\it Acknowledgements:}\ }
\newcount\eqcount
\eqcount=0
\def\Eqn#1{\global\advance\eqcount by 1
\ifnum\the\sectioncount=0
	\xdef#1{{\noexpand\oldsize\the\eqcount}}
	\eqno({\oldstyle\the\eqcount})
\else
        \ifnum\the\appendixcount=0
\xdef#1{{\noexpand\oldsize\the\sectioncount}.{\noexpand\oldsize\the\eqcount}}
                \eqno({\oldstyle\the\sectioncount}.{\oldstyle\the\eqcount})\fi
        \ifnum\the\appendixcount=1
	        \xdef#1{{\noexpand\oldstyle A}.{\noexpand\oldstyle\the\eqcount}}
                \eqno({\oldstyle A}.{\oldstyle\the\eqcount})\fi
        \ifnum\the\appendixcount=2
	        \xdef#1{{\noexpand\oldstyle B}.{\noexpand\oldstyle\the\eqcount}}
                \eqno({\oldstyle B}.{\oldstyle\the\eqcount})\fi
        \ifnum\the\appendixcount=3
	        \xdef#1{{\noexpand\oldstyle C}.{\noexpand\oldstyle\the\eqcount}}
                \eqno({\oldstyle C}.{\oldstyle\the\eqcount})\fi
\fi}
\def\eqn{\global\advance\eqcount by 1
\ifnum\the\sectioncount=0
	\eqno({\oldstyle\the\eqcount})
\else
        \ifnum\the\appendixcount=0
                \eqno({\oldstyle\the\sectioncount}.{\oldstyle\the\eqcount})\fi
        \ifnum\the\appendixcount=1
                \eqno({\oldstyle A}.{\oldstyle\the\eqcount})\fi
        \ifnum\the\appendixcount=2
                \eqno({\oldstyle B}.{\oldstyle\the\eqcount})\fi
        \ifnum\the\appendixcount=3
                \eqno({\oldstyle C}.{\oldstyle\the\eqcount})\fi
\fi}
\def\multi{\global\advance\eqcount by 1}
\def\multieq#1#2{\xdef#1{{\old\the\eqcount#2}}
        \eqno{({\oldstyle\the\eqcount#2})}}
\newtoks\url
\def\Href#1#2{\catcode`\#=12\url={#1}\catcode`\#=\active#2}
\def\href#1#2{{#2}}

\parskip=3.5pt plus .3pt minus .3pt
\baselineskip=14pt plus .1pt minus .05pt
\lineskip=.5pt plus .05pt minus .05pt
\lineskiplimit=.5pt
\abovedisplayskip=18pt plus 4pt minus 2pt
\belowdisplayskip=\abovedisplayskip
\hsize=14cm
\vsize=19cm
\hoffset=1.5cm
\voffset=1.8cm
\frenchspacing
\footline={}
\raggedbottom

\newskip\origparindent
\origparindent=\parindent

\def\ss{\scriptstyle}

\def\*{\partial}
\def\punkt{\,\,.}
\def\komma{\,\,,}

\def\={\!=\!}
\def\small#1{{\hbox{$#1$}}}

\def\fraction#1{\small{1\over#1}}
\def\fr{\fraction}
\def\Fraction#1#2{\small{#1\over#2}}
\def\Fr{\Fraction}
\def\tr{\hbox{\rm tr}}
\def\eg{{\it e.g.}}

\def\ie{{\it i.e.}}

\def\nlni{\hfill\break}

\def\a{\alpha}
\def\b{\beta}
\def\d{\delta}

\def\g{\gamma}
\def\l{\lambda}
\def\o{\omega}

\def\L{\Lambda}
\def\O{\Omega}

\def\w{\!\wedge\!}

\def\Int{\int\limits}


\def\L{\Lambda}
\def\o{\omega}
\def\w{\wedge}
\def\i{\imath}
\def\O{\Omega}

\def\lb{\bar\l}
\def\LB{\bar\L}
\def\ellb{\bar\ell}
\def\zb{\bar z}

\def\ol{\overline}
\def\ol{\bar}

\ref\ABC{Y. Aisaka and M. Cederwall, unpublished.}

\ref\BerkovitsNonMinimal{N. Berkovits,
{\xit ``Pure spinor formalism as an N=2 topological string''},
\jhep{05}{10}{2005}{089} [\hepth{0509120}].}

\ref\BerkovitsVanishing{N. Berkovits, {\xit ``Multiloop amplitudes and
vanishing theorems using the pure spinor formalism for the
superstring''}, \jhep{04}{09}{2004}{047} [\hepth{0406055}].}

\ref\BerkovitsNekrasovMultiloop{N. Berkovits and N. Nekrasov, {\xit
    ``Multiloop superstring amplitudes from non-minimal pure spinor
    formalism''}, \jhep{06}{12}{2006}{029} [\hepth{0609012}].}

\ref\MovshevEleven{M.V. Movshev,	
{\xit ``Geometry of a desingularization of eleven-dimensional
gravitational spinors''}, \hfill\break\arxiv{1105}{0127}.}

\ref\PureSGI{M. Cederwall, {\xit ``Towards a manifestly supersymmetric
    action for D=11 supergravity''}, \jhep{10}{01}{2010}{117}
    [\arxiv{0912}{1814}].}  

\ref\PureSGII{M. Cederwall, 
{\xit ``D=11 supergravity with manifest supersymmetry''},
    \MPLA{25}{2010}{3201} [\arxiv{1001}{0112}].}

\ref\GrassiGuttenberg{P.A. Grassi and S. Guttenberg, 
{\xit ``On projections to the pure spinor space''},
\arxiv{1109}{2848}.}

\ref\GriffithsHarris{P. Griffiths and J. Harris, {\xit ``Principles of
algebraic geometry''}, Wiley \&\ Sons (1978).}

\ref\CederwallNilssonTsimpisI{M. Cederwall, B.E.W. Nilsson and D. Tsimpis,
{\xit ``The structure of maximally supersymmetric super-Yang--Mills
theory --- constraining higher order corrections''},
\jhep{01}{06}{2001}{034} 
[\hepth{0102009}].}

\ref\CederwallNilssonTsimpisII{M. Cederwall, B.E.W. Nilsson and D. Tsimpis,
{\xit ``D=10 super-Yang--Mills at $\ss O(\a'^2)$''},
\JHEP{01}{07}{2001}{042} [\hepth{0104236}].}

\ref\BerkovitsParticle{N. Berkovits, {\xit ``Covariant quantization of
the superparticle using pure spinors''}, \jhep{01}{09}{2001}{016}
[\hepth{0105050}].}

\ref\SpinorialCohomology{M. Cederwall, B.E.W. Nilsson and D. Tsimpis,
{\xit ``Spinorial cohomology and maximally supersymmetric theories''},
\jhep{02}{02}{2002}{009} [\hepth{0110069}];
M. Cederwall, {\xit ``Superspace methods in string theory, supergravity and gauge theory''}, Lectures at the XXXVII Winter School in Theoretical Physics ``New Developments in Fundamental Interactions Theories'',  Karpacz, Poland,  Feb. 6-15, 2001, \hepth{0105176}.}

\ref\Movshev{M. Movshev and A. Schwarz, {\xit ``On maximally
supersymmetric Yang--Mills theories''}, \NPB{681}{2004}{324}
[\hepth{0311132}].}

\ref\BerkovitsI{N. Berkovits,
{\xit ``Super-Poincar\'e covariant quantization of the superstring''},
\jhep{00}{04}{2000}{018} [\hepth{0001035}].}

\ref\CederwallNilssonSix{M. Cederwall and B.E.W. Nilsson, {\xit ``Pure
spinors and D=6 super-Yang--Mills''}, \arxiv{0801}{1428}.}

\ref\CGNN{M. Cederwall, U. Gran, M. Nielsen and B.E.W. Nilsson,
{\xit ``Manifestly supersymmetric M-theory''},
\JHEP{00}{10}{2000}{041} [\hepth{0007035}];
{\xit ``Generalised 11-dimensional supergravity''}, \hepth{0010042}.
}

\ref\CGNT{M. Cederwall, U. Gran, B.E.W. Nilsson and D. Tsimpis,
{\xit ``Supersymmetric corrections to eleven-dimen\-sional supergravity''},
\jhep{05}{05}{2005}{052} [\hepth{0409107}].}

\ref\NilssonPure{B.E.W.~Nilsson,
{\xit ``Pure spinors as auxiliary fields in the ten-dimensional
supersymmetric Yang--Mills theory''},
\CQG3{1986}{{\xrm L}41}.}

\ref\HowePureI{P.S. Howe, {\xit ``Pure spinor lines in superspace and
ten-dimensional supersymmetric theories''}, \PLB{258}{1991}{141}.}

\ref\HowePureII{P.S. Howe, {\xit ``Pure spinors, function superspaces
and supergravity theories in ten and eleven dimensions''},
\PLB{273}{1991}{90}.} 



\ref\CederwallThreeConf{M. Cederwall, {\xit ``N=8 superfield formulation of
the Bagger--Lambert--Gustavsson model''}, \jhep{08}{09}{2008}{116}
[\arxiv{0808}{3242}]; {\xit ``Superfield actions for N=8 
and N=6 conformal theories in three dimensions''},
\jhep{08}{10}{2008}{70}
[\arxiv{0808}{3242}]; {\xit ``Pure spinor superfields,
with application to D=3 conformal models''}, \arxiv{0906}{5490}.}

\ref\BerkovitsICTP{N. Berkovits, {\xit ``ICTP lectures on covariant
quantization of the superstring''}, proceedings of the ICTP Spring
School on Superstrings and Related Matters, Trieste, Italy, 2002
[\hepth{0209059}.]} 

\ref\BatalinVilkovisky{I.A. Batalin and G.I. Vilkovisky, {\xit ``Gauge
algebra and quantization''}, \PLB{102}{1981}{27}.}

\ref\FusterBVReview{A. Fuster, M. Henneaux and A. Maas, {\xit
``BRST-antifield quantization: a short review''},\nlni\hepth{0506098}.}

\ref\BerkovitsMembrane{N. Berkovits,
	{\xit ``Towards covariant quantization of the supermembrane''},
	\JHEP{02}{09}{2002}{051} [\hepth{0201151}].}

\ref\BerkovitsNekrasovCharacter{N. Berkovits and N. Nekrasov, {\xit
    ``The character of pure spinors''}, \LMP{74}{2005}{75}
  \nlni[\hepth{0503075}].}

\ref\AnguelovaGrassiVanhove{L. Anguelova, P.A. Grassi and P. Vanhove,
  {\xit ``Covariant one-loop amplitudes in D=11''},
  \NPB{702}{2004}{269} [\hepth{0408171}].}

\ref\GrassiVanhove{P.A. Grassi and P. Vanhove, {\xit ``Topological M
    theory from pure spinor formalism''}, \ATMP{9}{2005}{285}
  [\hepth{0411167}].} 

\ref\BerkovitsNekrasovMultiloop{N. Berkovits and N. Nekrasov, {\xit
    ``Multiloop superstring amplitudes from non-minimal pure spinor
    formalism''}, \jhep{06}{12}{2006}{029} [\hepth{0609012}].}

\ref\BerkovitsII{N. Berkovits and B.C. Valillo, 
{\xit ``Consistency of super-Poincar\'e covariant superstring tree
amplitudes''}, \jhep{00}{07}{2000}{015} [\hepth{0004171}].}

\ref\BjornssonGreen{J. Bj\"ornsson and M.B. Green, {\xit ``5 loops in
25/4 dimensions''}, \jhep{10}{08}{2010}{132} [\arxiv{1004}{2692}].}

\ref\BjornssonMultiLoop{J. Bj\"ornsson, {\xit ``Multi-loop amplitudes
in maximally supersymmetric pure spinor field
theory''}, \jhep{11}{01}{2011}{002} [\arxiv{1009}{5906}].}

\ref\BerkovitsCherkis{N. Berkovits and S.A. Cherkis, {\xit
``Higher-dimensional twistor transforms using pure spinors''}, 
\jhep{04}{12}{2004}{049} [\hepth{0409243}].}

\ref\CederwallKarlsson{M. Cederwall and A. Karlsson, {\xit ``Pure
spinor superfields and Born--Infeld theory''}, \arxiv{1109}{0809}.}

\ref\GomezOneLoop{H. Gomez, {\xit ``One-loop superstring amplitude
from integrals on pure spinors space''}, \jhep{09}{12}{2009}{034}
[\arxiv{0910}{3405}].} 

\ref\NekrasovBetaGamma{N. Nekrasov, {\xit ``Lectures on curved
beta-gamma systems, pure spinors, and anomalies''}, \hepth{0511008}.}

\ref\AisakaArroyoBerkovitsNekrasov{Y. Aisaka, E.A. Arroyo,
N. Berkovits and N. Nekrasov, {\xit ``Pure spinor partition function
and the massive superstring spectrum''}, \jhep{08}{08}{2008}{050}
[arxiv{0806}{0584}].} 

\ref\AisakaArroyo{Y. Aisaka and E.A. Arroyo, {\xit ``Hilbert space of
curved $\ss\beta\gamma$ systems on quadric cones''},
\jhep{08}{08}{2008}{052} [\arxiv{0806}{0586}].}

\ref\AisakaOperatorSpace{Y. Aisaka, {\xit ``Operator space of pure
spinors''}, \PTPS{188}{2011}{227}.}


\headtext={M. Cederwall: 
``The geometry of pure spinor space''}

\line{
\epsfxsize=18mm
\epsffile{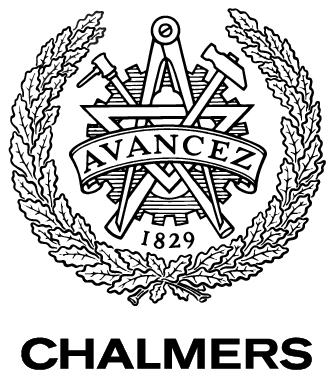}
\hfill}
\vskip-12mm
\line{\hfill Gothenburg preprint}
\line{\hfill November, {\old2011}}
\line{\hrulefill}

\vfill
\vskip.5cm

\centerline{\sixteenhelvbold
The geometry of pure spinor space} 

%

\vfill

\centerline{\twelvehelvbold{Martin Cederwall}}

\vfill
\vskip-1cm

\centerline{\xrm Fundamental Physics}
\centerline{\xrm Chalmers University of Technology}
\centerline{\xrm SE 412 96 Gothenburg, Sweden}

\vfill

{\narrower\noindent \underbar{Abstract:} We investigate the complex
geometry of $D=10$ pure spinor space. The K\"ahler structure and
the corresponding metric giving rise to the desired Calabi--Yau
property are determined, and an explicit covariant expression for
the Laplacian is given. The metric is not that of a c\^one obtained by
embedding pure spinor space in a flat space of unconstrained spinors. 
Some directions for future studies, concerning regularisation and 
generalisation to eleven dimensions, are briefly discussed.  
\smallskip}
\vfill

\font\xxtt=cmtt6

\vtop{\baselineskip=.6\baselineskip\xxtt
\line{\hrulefill}
\catcode`\@=11
\line{email: martin.cederwall@chalmers.se\hfill}
\catcode`\@=\active
}

\eject

\def\textfrac#1#2{\raise .45ex\hbox{\the\scriptfont0 #1}\nobreak\hskip-1pt/\hskip-1pt\hbox{\the\scriptfont0 #2}}

\section\Introduction{Introduction and background}As is well known, 
pure spinor superfield formalism solves the problem
of giving off-shell supersymmetric formulations of maximally
supersymmetric field theories and string theories
(see \eg\ refs. [\NilssonPure-\skipref\HowePureI\skipref\HowePureII\skipref\BerkovitsI\skipref\BerkovitsICTP\skipref\SpinorialCohomology\skipref\CederwallNilssonTsimpisI\skipref\CederwallNilssonTsimpisII\skipref\BerkovitsParticle\skipref\Movshev\skipref\CGNN\skipref\CGNT\skipref\CederwallNilssonSix\skipref\CederwallThreeConf\skipref\PureSGI\skipref\PureSGII\CederwallKarlsson]). It allows in
principle for the covariant calculation of amplitudes in such
theories, see refs.
[\BerkovitsNonMinimal-\skipref\BerkovitsII\skipref\BerkovitsNekrasovMultiloop\skipref\BjornssonGreen\BjornssonMultiLoop]
and references therein. 

Some problems remain, though. One of the main issues is how to make
integrals over the pure spinor space convergent. The pure spinor space
is non-compact and has disconnected boundary components at the origin
and at infinity. Divergences at infinity are easily dealt with 
[\BerkovitsNonMinimal]. Divergences at the origin are related to 
the question of gauge fixing. Normally, one constructs a gauge
fixing operator $b$ (the ``$b$ ghost'') as a differential operator and
demands that $b\Psi=0$ ($\Psi$ being the pure spinor superfield). This
operator [\BerkovitsNonMinimal], 
in its simplest version, is singular at the origin $\l=0$ of
pure spinor space, and a sufficient number of propagators containing
$b$ will generate divergences at the origin (while at the same time
fermionic integrals become over-saturated). A schematic, but
technically quite complicated, recipe for the regularisation of such
divergences at higher loops was given in
ref. [\BerkovitsNekrasovMultiloop]. This method demonstrates in
principle the existence of a regular $b$ operator, but can hardly be
said to provide an explicit form suited for calculations.

We are convinced that the work with trying to resolve these issues
will benefit from a clearer understanding of the geometry of the pure
spinor space, and this is the purpose of the present paper. The
questions above concerning gauge fixing and regularity will not be
solved here, but we hope this can provide a starting point. We comment on
this in the final section.

Many of the statements made here concerning pure spinor geometry are
already well known, but are included for sake of completeness. See
\eg\ refs. 
[\BerkovitsNonMinimal,\BerkovitsNekrasovMultiloop,\NekrasovBetaGamma-\skipref\AisakaArroyoBerkovitsNekrasov\skipref\AisakaArroyo\skipref\AisakaOperatorSpace\GomezOneLoop] 
for some geometrical considerations. The
main results of the present paper is the form of the K\"ahler
potential and the metric leading to the (necessary) Calabi--Yau
property, together with the covariant construction of metric-dependent
differential operators, such as the Laplacian.

\section\PureSpinorSpace{Properties of pure spinor space}The
pure spinor
constraint $(\l\g^a\l)=0$ determines an 11-dimensional complex space,
holomorphically embedded in the 16-dimensional space of unconstrained
spinors. In the non-minimal formalism [\BerkovitsNonMinimal], 
where integration is well
defined, one also considers the complex conjugate spinor $\lb_\a$ (of
course fulfilling $(\lb\g^a\lb)=0$) and
the fermionic variable $r_\a$, with $(\lb\g^ar)=0$. Due to the latter
constraint, the variables $r_\a$
may be identified with $d\lb_\a$. The non-minimal BRST operator is 
$$
Q=(\l D)+(r{\*\over\*\lb})=(\l D)+\ol\*\komma\eqn
$$
where $\ol\*=d\lb_\a{\*\over\*\lb_\a}$ is the usual Dolbeault
operator [\BerkovitsNekrasovMultiloop] and $D_\a$ the fermionic
covariant derivative.

Letting a field $\Psi$ depend also on $\lb$ and $r$ means letting it
be a cochain with antiholomorphic indices. An action may be written as
$$
S=\Int\Omega\wedge(\Psi\wedge Q\Psi+\ldots)\Eqn\FormAction
$$
(where integration over superspace coordinates $x$ and $\theta$ has
been suppressed for brevity). $\O$ here denotes a holomorphic top
form, which is described in section {\old2}.{\old2}.

\subsection\Coordinates{Coordinates on pure spinor space}The stability
group for a pure spinor is $SU(5)$. Spinors of the two chiralities,
${\bf16}$ and ${\bf\overline{16}}$ decompose under $SU(5)\times
U(1)\subset Spin(16)$ as 
$$
\eqalign{
{\bf16}&\rightarrow
{\bf1}_{-5/2}\oplus{\bf10}_{-1/2}\oplus{\bf\ol5}_{3/2}\komma\cr
{\bf\overline{16}}&\rightarrow
{\bf5}_{-3/2}\oplus{\bf\overline{10}}_{1/2}\oplus{\bf1}_{5/2}\punkt\cr
}\eqn
$$
Consider a chiral spinor $\L$ in ${\bf16}$ as an even form, 
$$
\L=\L_0+\L_2+\L_4\punkt\Eqn\SUFiveBasis
$$
We will write $\ell\equiv\L_0$. 
The $\g$-matrices act as 
$\o\w\L$ and $\i_v\L$, where $\o$ is a 1-form and $v$ a vector, and
this extends also to the odd forms, containing a spinor of opposite
chirality.
The $so(10)$ generators outside $su(5)\oplus u(1)$ are given as
$\mu\w$ and $\i_m$, where $\mu$ is a 2-form and $m$ an antisymmetric bivector.
The invariant product of ${\bf16}$ and ${\bf\overline{16}}$
is 
$$
<\L,K>\,={\star}(\L_0K_5-\L_2\w K_3+\L_4\w K_1)\punkt\eqn
$$

The pure spinor constraint $(\l\g^a\l)=0$ becomes
$$
<\L,\o\w\L>\,=0=\,<\L,\i_v\L>\punkt\eqn
$$
The first of these conditions imply that
$\L_4=\fr2\ell^{-1}\L_2\w\L_2$ (in the patch where $\ell\neq0$). 
Once it is satisfied, 
$<\L,\i_v\L>\sim\ell^{-1}{\star}(\i_v\L_2\w\L_2\w\L_2)=0$.
The 11-dimensional space of $D=10$ pure spinors is parametrised by $\ell$ and
$\L_2$.

\subsection\KahlerCY{K\"ahler structure and the holomorphic
top form}There is an 
$so(10)$-invariant $(11,0)$-form [\BerkovitsNekrasovMultiloop] on pure
spinor space,
$$
\O=\ell^{-3}d\l \,d^{10}\L_2\punkt\Eqn\TenVolumeForm
$$
It is clear from above that its $U(1)$ charge is $0$, and it remains
to show the invariance under $\mu\w$ and $\i_m$ as above. The first of
these is trivial, since it does not pull down any non-linear
contribution from $\L_4$. Under $\i_m$ we have
$$
\eqalign{
\d_m\ell&=\i_m\L_2\komma\cr
\d_m\L_2&=\i_m\L_4=\fr2\ell^{-1}\i_m(\L_2\w\L_2)\punkt\cr
}\eqn
$$
Using the second equation, together with
$d\L_{AB}(d^9\L)^{CD}=\fr{10}\d_{AB}^{CD}d^{10}\L_2$ where
$(d^9\L_2)^{AB}$ is defined by
$d^{10}\L_2\equiv d\L_{AB}(d^9\L)^{AB}$, a short
calculation shows that the volume form (\TenVolumeForm) is invariant
under $so(10)$ precisely when the power of $\ell$ in the prefactor is $-3$.
The factor $\ell^{-3}$ of course reflects the same power of $\l$ in the
covariant ``measure'', given as
$$
\O\propto(\l\lb)^{-3}\lb_{\a_1}\lb_{\a_2}\lb_{\a_3}
T^{\a_1\a_2\a_3}{}_{\b_1\ldots\b_{11}}d\l^{\b_1}\wedge\ldots
d\l^{\b_{11}}\komma\Eqn\CovariantTopForm
$$
$T$ being the unique $so(10)$-invariant tensor with 3 pure spinor
indices and 11 antisymmetric cospinor indices (\ie, Clebsch--Gordan
coefficients for the formation of a singlet from 
$(00030)\otimes(\wedge^{11}(00001)$). It was demonstrated in
ref. [\BerkovitsCherkis] that $\O$ of eq. (\CovariantTopForm) is
independent of $\lb$, \ie, that $\ol\*\O=0$. 
The $so(10)$ invariance ensures that $\O$ of eq. (\TenVolumeForm) 
is globally defined. The existence of
such an $\Omega$ is the Calabi--Yau property.

The existence of the holomorphic top form 
is essential for the
gauge invariance of the action (\FormAction), since $Q$ can not be
partially integrated if not $\ol\*\O=0$. Equally important is
the (obvious) non-exactness of $\O$, $\O\neq\ol\*\xi$.
In the field $\Psi$, all cohomology has minimal representatives independent of
$\lb$ and $r$, \ie, holomorphic $(0,0)$-forms. For the action not to
produce an unwanted doubling of components fields, the
anti-holomorphic top form $\ol\O$ must not represent $\bar\*$ cohomology,
and thus $\ol\O=\bar\*\xi_{(0,10)}$. Such a form $\xi_{(0,10)}$ is
readily constructed as
$$
\xi_{(0,10)}\propto(\l\lb)^{-3}\l^{\a_1}\l^{\a_2}\l^{\a_3}
\bar T_{\a_1\a_2\a_3}{}^{\b_1\ldots\b_{11}}\lb_{\b_1}d\lb_{\b_2}\wedge\ldots
d\lb_{\b_{11}}\punkt\eqn
$$
This asymmetry in the $\ol\*$ cohomology is possible since the
pure spinor space is non-compact. Also the volume form 
$\hbox{Vol}=\O\wedge\ol\O$ is cohomologically trivial. This does
not mean that the ``volume'' (which needs regularisation at infinity)
vanishes, since there are two boundary components, at zero and infinity.
The K\"ahler potential is globally defined, so neither does the K\"ahler form 
represent cohomology, which is consistent with the triviality of the
volume form.

The full complex
$16_{{\Bbb C}}$-dimensional spinor space allows a K\"ahler structure
(in fact, infinitely many). 
The flat
geometry corresponds to the K\"ahler potential $K_0=(\l\lb)$. Contrary
to what is sometimes assumed, this
will {\it not} be the actual geometry, but we will
nevertheless examine
the induced geometry on the pure spinors. We use the coordinates
$(z^m;\zb^{\ol m})=(\ell,\L_{ab};\ellb,\ol\L^{ab})$ from
above. 
The K\"ahler
potential for the induced geometry is inherited from the embedding space:
$$
K_0=(\l\lb)=\,<\L,\LB>\,=\ell\ellb-\fr2\tr(\L\LB)
-\fr4(\ell\ellb)^{-1}X\komma\Eqn\FlatKahler
$$ 
where $X=\tr(\L\LB)^2-\fr2\left(\tr(\L\LB)\right)^2$,
and the K\"ahler form is given as 
$\o_0=h_{m\ol n}dz^m\wedge d\zb^{\ol n}=\*\ol\*K_0$ 
(we use ``$h$'' for this metric and reserve ``$g$'' for later). 
The explicit form of the metric is explored in the following subsection.
The volume form is 
$\hbox{Vol}_0=\sqrt{h}\,d^{11}zd^{11}\zb\propto\wedge^{11}\o_0$.
Now, since $K_0$ is homogeneous of degree $(1,1)$ in the coordinates, it
is obvious that the metric, and thus also its determinant, is
homogeneous of degree $(0,0)$. A calculation using Mathematica shows
that indeed
$$
\sqrt{h}={(\l\lb)^3\over(\ell\ellb)^3}\punkt\Eqn\FlatDeterminant
$$
This is not acceptable --- as argued in the previous subsection 
it is crucial that there is a holomorphic
$(11,0)$-form $\O$ such that $\hbox{Vol}\propto\O\wedge\ol\O$, \ie,
that $\sqrt g$ factorises as $\sqrt g=f\ol f$, where $f(z)$ is
holomorphic.  

The only Lorentz invariant that does not vanish on the pure spinor
space is $(\l\lb)$. Therefore, a K\"ahler potential must be taken as a
function of this invariant. If we take $K=
\Fr{11}8K_0^{8/11}=\Fr{11}8(\l\lb)^{8/11}$, 
the metric will be homogeneous in both $z$
and $\ol z$ of degree $-{3\over11}$, and $\sqrt g$ will be of degree
$-3$.
Again, using Mathematica, we see that $\sqrt g=
\Fr8{11}(\ell\ellb)^{-3}$, so the invariant holomorphic volume form
(\TenVolumeForm) is
reproduced, and $\hbox{Vol}\propto\wedge^{11}\o\propto\O\wedge\ol\O$.
More generally, a K\"ahler potential $K_p=p^{-1}(\l\lb)^p$ gives
$\sqrt{g_p}=p(\l\lb)^{11p-8}(\ell\ellb)^{-3}$.

So the metric on the full spinor space giving the
induced metric is
$$
ds^2=
(\l\lb)^{-3/11}\left(d\l d\lb-\Fr3{11}(\l\lb)^{-1}(d\l\lb)(\l
d\lb)\right)
\komma\Eqn\InducedCovMetric
$$
obtained from the K\"ahler potential $K=\Fr{11}8(\l\lb)^{8/11}$.
Pure spinor space should not be thought of as a c\^one embedded in
flat space, but in a space with a metric which is already highly
singular at the origin.

This makes the pure spinor space Ricci flat. On K\"ahler manifolds,
the Ricci form (related to the Ricci tensor the same way as the K\"ahler
form to the metric) is given by
$\varrho=\*\ol\*\log\sqrt g$,
which vanishes when $\sqrt g$ factorises as $\sqrt g=f\ol f$, as
above. With
$K_p=p^{-1}(\l\lb)^p$, one gets
$\varrho=(11p-8)\*\ol\*\log(\l\lb)$. This vanishes only when
$p=\Fr8{11}$ and has rank $10$ for other values of $p$ (its components
along $\l$ and $\lb$ vanish). The scalar curvature is
$R=20(11p-8)(\l\lb)^{-1}$.



\subsection\ExplMetric{The explicit metric}The metric in the system 
with coordinates $(\ell,\L_{AB})$ defined
above can be given by using the pure spinor constraint in the
expression for the metric (or, the K\"ahler form) of the embedding
space. Let us first do this for the flat embedding space metric with 
K\"ahler potential $K_0=(\l\lb)$. We get
$$
\eqalign{
h^{0\ol0}&=1-\fr4(\ell\ellb)^{-2}X\komma\cr
h^{AB,0}&=-(\ell\ellb)^{-2}\ell
          \left[(\LB\L\LB)^{AB}-\fr2\LB^{AB}\tr(\L\LB)\right]\komma\cr
h^{AB,\ol C\ol D}
&=2\left[1-\fr2(\ell\ellb)^{-1}\tr(\L\LB)\right]\d^{AB}_{CD}
    +(\ell\ellb)^{-1}
\left[\LB^{AB}\L_{CD}+4\d_{[C}{}^{[A}(\L\LB)_{D]}{}^{B]}\right]\punkt\cr
}\Eqn\FlatMetric
$$

It is difficult to invert this metric directly. Instead we make use of
our knowledge of the embedding space metric in the
following procedure. Any tangent space vector can be thought of
$so(10)$-covariantly as being projected by the projection 
operator\foot\dagger{This
projection also occurs in ref. [\GrassiGuttenberg]. With $\lb$
replaced by a constant pure spinor, such operators appear also \eg\ in
ref. [\BerkovitsNekrasovMultiloop].} 
$$
P^\a{}_\b=\d^\a{}_\b-\fr2(\l\lb)^{-1}(\g^a\lb)^\a(\g_a\l)_\b
=(\l\lb)^{-1}\left[-\fr4\l^\a\lb_\b+\fr8(\g^{ab}\l)^\a(\g_{ab}\lb)_\b\right]
\komma\eqn
$$ 
acting
as the identity on any spinor $v^\a$ with $(v\g^a\l)=0$. A
conjugate tangent vector is projected by $\ol P=P^t$. When the
embedding space metric is the flat one, $P^{\a\ol\b}$ can be thought
of as the metric. Its inverse on the tangent directions is the metric
$P$ itself. The inverse metric in some coordinate system is obtained
from this inverse metric. Specifically, in the $SU(5)$-covariant
coordinate system, we take cotangent vectors decomposed as in
eq. (\SUFiveBasis), but without 4-form (or 1-form): 
$V=V_3+V_5$, $\ol V=\ol V_0+\ol V_2$. Contracting these cotangent
vectors with the covariant inverse metric gives the inverse metric in
the coordinate system, which reads explicitly:
$$
\eqalign{
(h^{-1})_{0\ol0}&=1+\fr4(\l\lb)^{-1}(\ell\ellb)^{-1}X\komma\cr
(h^{-1})_{AB,\ol0}&=(\l\lb)^{-1}(\ell\ellb)^{-1}\ellb
     \left[(\LB\L\LB)_{AB}-\fr2\L_{AB}\tr(\L\LB)\right]\komma\cr
(h^{-1})_{AB,\ol C\ol D}
&=2\left[1+\fr2(\l\lb)^{-1}\tr(\L\LB)
          +\fr2(\l\lb)^{-1}(\ell\ellb)^{-1}X\right]\d_{AB}^{CD}\cr
&-(\l\lb)^{-1}\left[\L_{AB}\LB^{CD}
           +4\d_{[A}{}^{[C}(\L\LB)_{B]}{}^{D]}\right]\cr
&-4(\l\lb)^{-1}(\ell\ellb)^{-1}\d_{[A}{}^{[C}
    \left[(\L\LB\L\LB)_{B]}{}^{D]}
      -\fr2(\L\LB)_{B]}{}^{D]}\tr(\L\LB)\right]
            \punkt\cr
}\Eqn\InverseFlatMetric
$$
Already in this expression we have used some identities from the
Appendix to simplify the expressions. The expression $(\l\lb)$ should
here always be understood as the expression in eq. (\FlatKahler).
We have verified that the metric and inverse metric of eqs. 
(\FlatMetric) and (\InverseFlatMetric) satisfy $hh^{-1}=1$, but this
is a lengthy calculation involving the relations in the Appendix.

We now turn to the actual metric on the pure spinor space allowing for
the holomorphic volume form $\Omega$. The covariant form
(\InducedCovMetric) can be understood as a metric
$$
G_{\a\ol\b}=(\l\lb)^{-3/11}\left[P_{\a\ol\b}
-\Fr3{11}(\l\lb)^{-1}\lb_\a\l_{\ol\b}\right]\punkt\eqn
$$ 
The last term is automatically tangent. Its inverse $\tilde G$ 
on tangent space (by which we mean $G\tilde G=P$) is
$$
\tilde G^{\a\ol\b}=(\l\lb)^{3/11}\left[P^{\a\ol\b}
      +\Fr38(\l\lb)^{-1}\l^\a\lb^{\ol\b}\right]
=\fr8(\l\lb)^{-8/11}\left[\l^\a\lb^{\ol\b}
     +(\g^{ab}\l)^\a(\g_{ab}\lb)^{\ol\b}\right]
\punkt\eqn
$$
In the $SU(5)$ coordinate system, using the same procedure as before
to form the metric and its inverse, this amounts to
$$
\eqalign{
g&=(\l\lb)^{-3/11}\left[h
   -\Fr3{11}(\l\lb)^{-1}(\b\otimes\ol\b+\ol\b\otimes\b)\right]\komma\cr
g^{-1}&=(\l\lb)^{3/11}\left[h^{-1}
   +\Fr38(\l\lb)^{-1}(\g\otimes\ol\g+\ol\g\otimes\g)\right]\komma\cr
}\eqn
$$ 
where
$$
\eqalign{
\b^0&=\left[1+\fr4(\ell\ellb)^{-2}X\right]\ellb\komma\cr
\b^{AB}&=\LB^{AB}
   +(\ell\ellb)^{-1}\left[(\LB\L\LB)^{AB}-\fr2\LB^{AB}\tr(\L\LB)\right]
         \komma\cr
\g_0&=\ell\komma\cr
\g_{AB}&=\L_{AB}\punkt\cr
}\eqn
$$
It is easily checked that $\ol\b=h\g$ and that
$\b\cdot\g=(\l\lb)$, ensuring that $g^{-1}$ is the inverse of $g$.

\subsection\CovariantExpr{Covariant expressions for operators}The
inverse metric is needed for
the construction of metric dependent differential operators, 
\eg\ the Laplacian. The Laplacian on a K\"ahler space
simplifies, in that it can be given in terms of the Dolbeault operator
$\ol\*$ as $\Delta_{\ol\*}=\{\ol\*,\ol\*^\star\}$, where
$\ol\*^\star=\star\ol\*\star$ is the adjoint operator to 
$\ol\*$. $\Delta_{\ol\*}$ is
proportional to the ordinary Laplacian, 
$2\Delta_{\ol\*}=\Delta_d=\{d,d^\star\}$. 

In ref. [\BerkovitsNekrasovMultiloop], regularisation of higher loop
integrals was performed by using an operator that was argued to have
some close relationship with the Laplacian (although we suspect that
if there is an identity it may well be with the Laplacian
corresponding to the metric $h$). The $b$ operator was shown to become
regular when modified as 
$b'=e^{t\{Q,\chi\}}b\,e^{-t\{Q,\chi\}},$
with the "regulating fermion" $\chi$, roughly speaking, being proportional to
$\ol\*^\star$. 

We have
$\{Q,\ol\*^\star\}=\Delta_{\ol\*}+\ldots$.
The operator $\ol\*^\star$ should be formed as the
divergence $g^{m\ol n}\ol\imath_{\ol n}\*_m$. Here the contraction
may be represented as a field $s$ and the derivative as a $w$. In this
picture, these spinor operators must come in some gauge invariant
combination. It is clear from above how this is achieved. One should
use the $so(10)$-covariant form $\tilde G$ of the inverse metric:
$$
\ol\*^\star=\tilde G^{\a\ol\b}s_{\ol\b}w_\a\punkt\eqn
$$
Consequently, 
$$
\{Q,\ol\*^\star\}=\Delta_{\ol\*}-\tilde G^{\a\ol\b}s_{\ol\b}D_\a\punkt\eqn
$$
This also gives an $so(10)$-covariant prescription for the
Laplacian. As an example, the Laplacian acting on a $(0,0)$-form
$\phi$ is
$$
\Delta_{\ol\*}\phi=\ol\*^\star\ol\*\phi
=\tilde G^{\ol\b\a}w_\a\ol w_{\ol\b}\phi
=\fr8(\l\lb)^{-8/11}(N\ol N+N^{ab}\ol N_{ab})\phi\komma\eqn
$$
where $N$ and $N^{ab}$ are the well defined quantities 
$N=(\l w)$, $N^{ab}=(\l\g^{ab}w)$.
In principle, one could think of other constructions, \eg\ using the
matrix $P$ instead of $\tilde G$. Such operators will however be less
natural, and will not have a geometric interpretation on pure spinor
space.

We have not yet checked whether or not such a geometric regularisation
will yield a regular $b$ operator, but find it plausible considering the
work in ref. [\BerkovitsNekrasovMultiloop]. We comment more on related
issues in the concluding section.

\vfill\eject

\section\FurtherDirections{Further directions}It would be interesting
if a more geometric approach can help in defining regular operators
(in particular, the $b$ operator) on pure spinor space. Any gauge fixing
operator $b$ should have the property $\{Q,b\}\propto p^2$, but this does
not uniquely fix $b$. Instead there is a gauge degree of freedom
amounting to shifting $b$ with something $Q$-exact, and this is the
freedom one uses in regularisation. One possibility might lie in using
the localisation obtained by letting the parameter $t$ in the exponent
tend to infinity, which would mean only dealing with zero eigenstates
of $\Delta_{\ol\*}-\tilde G^{\a\ol\b}s_{\ol\b}D_\a$. This would be a
quite natural $Q$-exact generalisation of choosing harmonic forms as
representatives for cohomology. We would like to examine whether this
or some similar geometrically motivated procedure can provide a good
regularisation. 

Another issue is generalisation to $D=11$. In order to investigate
properties of amplitudes in dimensional reductions of $D=11$
supergravity, we believe that the covariant action of
refs. [\PureSGI,\PureSGII] will be the best starting point. 
Some work has been done in a first-quantised formalism 
[\AnguelovaGrassiVanhove,\GrassiVanhove].
The
geometry of $D=11$ pure spinor space is more complicated than in
$D=10$ 
[\BerkovitsMembrane,\SpinorialCohomology,\PureSGI,\PureSGII,\MovshevEleven],
mainly due to the existence of two independent $so(11)$ 
scalars, $(\l\lb)$ and $(\l\g^{ab}\l)(\lb\g_{ab}\lb)$. The K\"ahler
potential will depend only on these.
The $b$ operator has been constructed [\ABC], but further regularisation will
certainly be needed.

\appendix{Some matrix identities}A Cayley--Hamilton 
relation for matrix products of $\L$'s and $\LB$'s, 
obtained from antisymmetrisation in 6 indices:
$$
0=\L\LB\L\LB\L-\fr2\L\LB\L\tr(\L\LB)
-\fr4\L X\komma\Eqn\CayleyHamiltonRel
$$
where $X=\tr(\L\LB)^2-\fr2\left(\tr(\L\LB)\right)^2$.
Another useful relation\foot\star{Note
that the existence of some such
relation can be deduced from the fact that the tensor product of the
symmetric tensor product of 2 $\L$'s (in $(0100)$) and the symmetric
tensor product of 2 $\LB$'s (in $(0010)$) only contains 3 structures
in $(0110)$, and this module is contained in the 4 terms of the first
line of the identity. Thus there must be $4-3=1$ identity.}, 
necessary for checking the form of the
inverse metric, is
$$
\eqalign{
0&=\L_{AB}(\LB\L\LB)^{CD}+(\L\LB\L)_{AB}\LB^{CD}
+2(\L\LB)_{[A}{}^C(\L\LB)_{B]}{}^D-\fr2\L_{AB}\LB^{CD}\tr(\L\LB)\cr
&+4\d_{[A}{}^{[C}\left[(\L\LB\L\LB)_{B]}{}^{D]}
       -\fr2(\L\LB)_{B]}{}^{D]}\tr(\L\LB)\right]
-\fr2\d_{AB}^{CD}X\komma\cr
}
\Eqn\FourLambdaRel
$$
which can also been obtained by cycling in 6 indices. It is
straightforward to check that it is consistent with contraction by
$\d_D^B$ (by direct calculation) and with $\L_{CD}$ or $\LB^{AB}$ (thanks to
eq. (\CayleyHamiltonRel)). As a consequence of eqs. (\FourLambdaRel) and
(\CayleyHamiltonRel) one
also gets 
$$
\eqalign{
0&=(\L\LB\L)_{AB}(\LB\L\LB)^{CD}+4(\L\LB)_{[A}{}^{[C}(\L\LB\L\LB)_{B]}{}^{D]}\cr
&-(\L\LB)_{[A}{}^{[C}(\L\LB)_{B]}{}^{D]}\tr(\L\LB)+\fr4\L_{AB}\LB^{CD}X
\punkt\cr
}\eqn
$$


\acknowledgements The author would like to thank M\aa ns Henningson for
generously sharing his knowledge on algebraic geometry.

\refout

\end